\documentclass[letters,fleqn,usenatbib]{mnras}

\usepackage[T1]{fontenc}
\usepackage{ae,aecompl}

\usepackage{graphicx}	
\usepackage{amsmath}	

\usepackage{amssymb}	

\def\ISMabs{{\tt ISMabs}}
\def\IGMabs{{\tt IGMabs}}

\def\Oi{O\,{\sc i}}
\def\Oii{O\,{\sc ii}}
\def\Oiii{O\,{\sc iii}}

\def\Ovi{O\,{\sc vi}}
\def\Ovii{O\,{\sc vii}}
\def\Oviii{O\,{\sc viii}}

\def\Cii{C\,{\sc ii}}

\def\NOi{{\it N}(\Oi)}
\def\NOii{{\it N}(\Oii)}
\def\NOiii{{\it N}(\Oiii)}
\def\NOvi{{\it N}(\Ovi)}
\def\NOvii{{\it N}(\Ovii)}
\def\NOviii{{\it N}(\Oviii)}

\title[OVI in the milky way]{On the discrepancy between the X-ray and UV absorption measurements of \Ovi\ in the local ISM}

\author[Gatuzz ]{
Efra\'in~Gatuzz$^{1,2}$\thanks{E-mail: efraingatuzz@gmail.com}
Javier~A.~Garc\'ia$^{3,4}$
and Timothy~R.~Kallman$^{5}$   
\\
$^{1}$ESO, Karl-Schwarzschild-Strasse 2, D-85748 Garching bei M\"unchen, Germany \\
$^{2}$Excellence Cluster Universe, Boltzmannstr. 2, D-85748, Garching, Germany\\
$^{3}$Cahill Center for Astronomy and Astrophysics, California Institute of Technology, Pasadena, CA 91125, USA\\
$^{4}$Dr. Karl Remeis-Observatory and Erlangen Centre for Astroparticle Physics, Universit\"at Erlangen-N\"urnberg, Sternwartstr. 7, \\96049 Bamberg, Germany\\
$^{5}$NASA Goddard Space Flight Center, Greenbelt, MD 20771, USA\\
} 
\date{Accepted XXX. Received YYY; in original form ZZZ}

\pubyear{2018}

\begin{document}
 \label{firstpage}
\pagerange{\pageref{firstpage}--\pageref{lastpage}}
\maketitle

\begin{abstract}
The total amount of \Ovi\ present in the interstellar medium
(ISM) obtained via absorption measurements in UV and X-ray
spectra is currently in disagreement, with the latter being significantly
larger (by a factor of 10 or more) than the former. Previous works have
proposed that the blend of the \Ovi\ K$\alpha$ line (22.032~\AA) with the \Oii\ K$\beta$-L12 line (22.04~\AA) could account for the stronger absorption observed in
the X-ray spectra. Here we present a detailed study of the oxygen absorption in
the local ISM, implementing our new model \IGMabs\, which includes
photoabsorption cross-sections of highly ionized species of abundant elements
as well as turbulence broadening.  By analyzing high-resolution {\it Chandra}
spectra of 13 low mass X-ray binaries (LMXBs) and 29 extragalactic sources, we
have estimated the column densities of \Oi--\Oiii\ and from \Ovi--\Oviii\ along
multiple line-of-sights. We find that in most cases the \Oii\ K$\beta$-L12 line accounts for < 30$\%$ of the total \Ovi\ K$\alpha$ + \Oii\ K$\beta$. We conclude that the amount of \Oii\ predicted by our model is still insufficient to
explain the discrepancy between X-ray and UV measurements of \Ovi\ column densities.
\end{abstract} 

\section{Introduction}\label{sec_in}
The interstellar medium (ISM) is one of the most important ingredients in the star life cycles, affecting the dynamics of the Galaxy. This complex environment can be analyzed by using astronomical observations from different energy bands, including optical \citep[e.g.][]{wel10,bra12,bai15,sch16}, radio \citep[e.g.][]{kal05,wil13,and15,mos17}, infrared \citep[e.g.][]{jac08,gar14,gia15,xue16}, UV \citep[e.g.][]{bow08,wak12,sav17} and X-rays \citep[e.g.][]{jue04,jue06,gat13a,mil15,nic16,gat18a,gat18b}. X-rays photons, in particular, are capable of trace multiple phases of the ISM, namely cold, warm and hot. Given that most of the baryonic matter resides among galaxies in the warm-hot phases \citep{shu12,mcq16}, X-ray spectra constitute a useful way to measure column densities, ionization state and abundances of the intergalactic medium \citep{nic16,fae17,arc18}.

 The hot component of the ISM, in particular, can be traced by \Ovi\ absorption lines in both, the ultra-violet (UV) and X-ray bands, providing a useful tool to study the abundance and ionization structure of the plasma \citep{pra00}. However, even when the integrated \Ovi\ column density measured with UV instruments must be equal to the obtained from the X-ray data, there is a well known mismatch between both measurements with differences as large as a factor of a few up to one order of magnitude \citep{ara03,mat17}. There are several contaminants that would affect the \NOvi\ measured by UV data. For example, the 1038 \AA\ feature is affected by \Cii$^{*}$ and H$_{2}$ absorption while the 1032 \AA\ feature is affected by H$_{2}$ and HD \citep{pat11}. However, when modeling in detail such contaminants the UV column densities become smaller, hence, creating a bigger gap between the X-ray and UV data \citep{sar17}.  Given that in UV the lines are resolved and unsaturated, the atomic parameters are well known,
and the resulting column densities make physical sense considering the mechanisms that could produce \Ovi, it is unlikely for UV measurements to be inaccurate.

  Recently, \citet{mat17} proposed that the mismatch between column densities is due to contamination
of the \Ovi\ K$\alpha$ absorption line, located at $22.032$ \AA\ \citep{mcl17}, with
the \Oii\ K$\beta$ ``line~12'' (L12, thereafter) transition located at $22.04$ \AA\ \citep{biz15}. By analyzing {\it XMM-Newton} spectra from 23 sources they measured equivalent widths (EWs) for the \Oii\ K$\alpha$ transition in order to estimate the \Oii\ K$\beta$-L12 level of contamination to the \Ovi\ K$\alpha$ absorption line. They concluded that the Galactic \NOvi\ measured in UV is too small to be detectable in X-ray and that the $22.032$ \AA\ X-ray absorption line is almost entirely dominated by the \Oii\ K$\beta$-L12 transition.  It should be noted that their conclusions are only valid for $z=0$ systems while discrepancies at higher redshift require other explanations.  We are testing this idea by performing a detailed analysis of the oxygen X-ray absorption in the local ISM along multiple line-of-sights, with emphasis on the \Ovi\ absorption
features, and the influence of the \Oii\ absorption on its modeling.  The
outline of this report is as follows. In Section~\ref{sec_dat} we explain the
data selection and data reduction processes. In Section~\ref{sec_results} we
describe and discuss the results obtained from the spectral fitting.  Finally,
Section~\ref{sec_con} summarizes our main conclusions.

\section{Observations and fitting procedure}\label{sec_dat}

In order to measure the \Ovi\ absorption in the local ISM we used a sample of both
galactic and extragalactic sources based on the sources previously analyzed by \citet{gat18a}. We
decided to exclude {\it XMM-Newton} observations due to to the presence of
instrumental features in the Reflection Grating Spectrometers
\citep[RGS;][]{den01}. These instrumental imperfections contaminate the \Ovi\
K$\alpha$  absorption feature at 22.032~\AA, which is of central interest in
this work. Thus, the final sample consists of 13 LMXBs and 29 extragalactic
sources.  Observations were reduced following the standard CIAO
threads\footnote{\url{http://cxc.harvard.edu/ciao/threads/gspec.html}} and the
zero-order position were estimated with the {\tt findzo}
algorithm\footnote{\url{http://space.mit.edu/cxc/analysis/findzo/}}. For each
source, all observations were combined using the {\tt
combine\_grating\_spectra} command. The spectral fitting was performed with the
{\sc xspec} software (version
12.9.1p\footnote{\url{https://heasarc.gsfc.nasa.gov/xanadu/xspec/}}).  Finally,
we used $\chi^{2}$ statistics in combination with the \citet{chu96} weighting
method which allows the analysis of low-counts spectra.

Gaussian profiles are commonly used in order to model X-ray absorption features
associated to the ISM. However, if the analysis is focused in only one
absorption line, ignoring the presence of absorption features due to other
transitions of the same ion or due to the presence of different ions, the
column densities can be over(under)-estimated \citep{gat14,gat16}.  To address
this issue, we have built a model for the entire oxygen K-shell region
(18--24~\AA), which includes absorption features from \Oi--\Oviii\ ions. In order to fit the absorption features from the low ionization ions
\Oi--\Oiii\, we use the \ISMabs\ model \citep{gat15}. These ions are associated to the neutral-warm components of the ISM
\citep{gat16,gat18a}. Then, we developed a new model called \IGMabs, which is similar to {
\ISMabs\ but includes photoabsorption
cross-sections from highly ionized ions, in order to model absorption features associated to \Ovi--\Oviii\ ions.

Figure~\ref{fig_O} shows the oxygen photoabsorption cross-sections included in the present analysis. Note that, in the computation of such cross-sections, the Auger broadening can be the dominant broadening process, and is included. As the plot shows, the \Ovi\ K$\alpha$ resonance lies close to the \Oii\ K$\beta$-L12 transition and the cross-section is larger for the former. It is important to note that the
\IGMabs\ model also includes a turbulence velocity parameter $v_{turb}$,
which is required when modeling the hot component of the ISM. Following the analysis in
\citet{gat18a}, we have fixed the $v_{turb}$ parameter of the \IGMabs\
model to $60$~km~s$^{-1}$ for the Galactic sources and to $110$~km~s$^{-1}$ for 
the Extragalactic sources.  Finally, for each source we have used a {\tt powerlaw} component to model the
continuum and we have fixed the $N({\rm H})$ of the \ISMabs\
model to the 21~cm values from \citet{wil13}. In this way, the free parameters
in our fitting procedure are the oxygen column densities and the {\tt powerlaw}
parameters.

  \begin{figure}
   \begin{center}
     \includegraphics[scale=0.37]{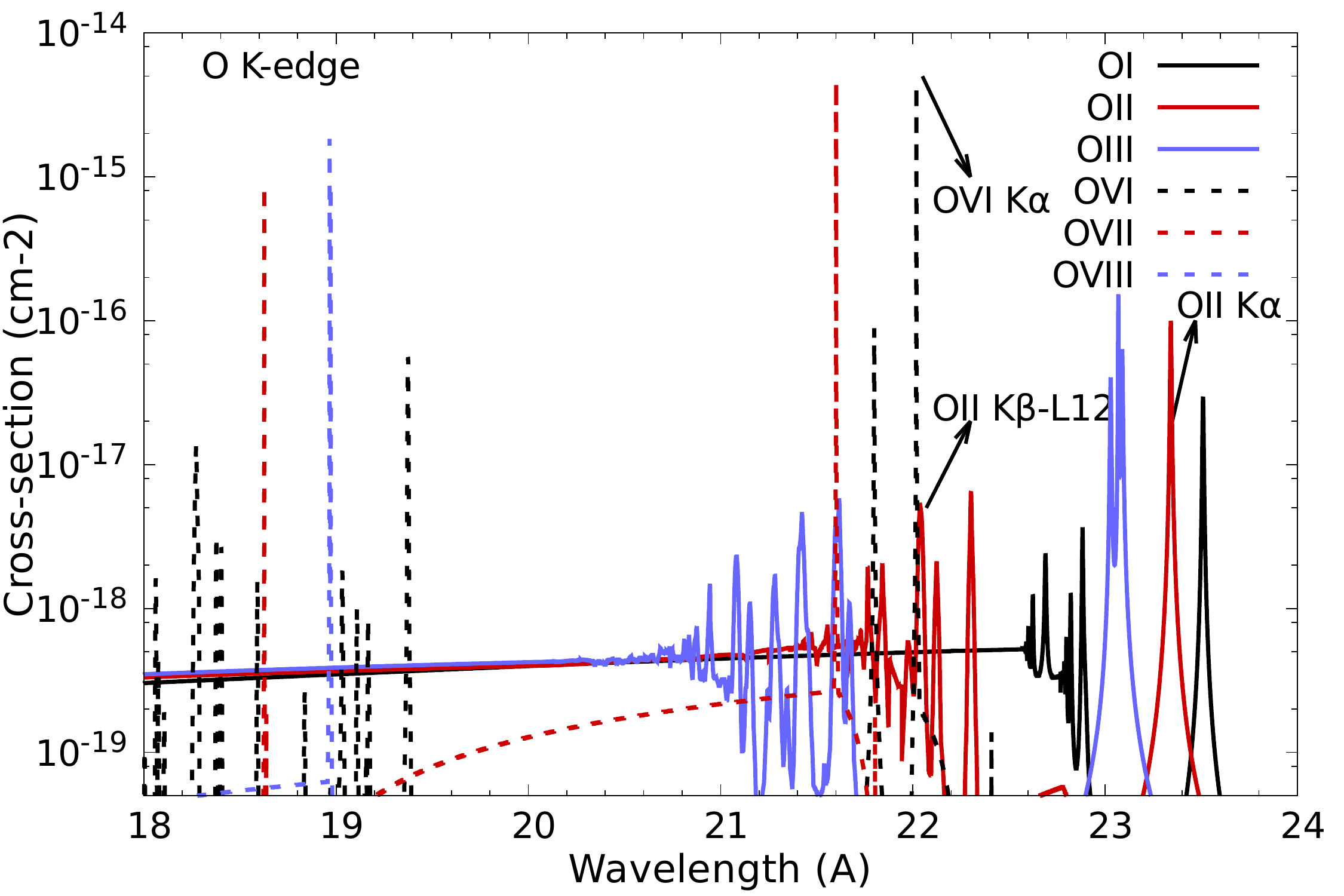}
          \vspace*{-6mm}
     \caption{ Oxygen photoabsorption cross sections that are implemented in the {\tt ISMabs} (\Oi--\Oiii\ ions) and {\tt IGMabs}  (\Ovi--\Oviii\ ions) model.}\label{fig_O}
     \end{center}
   \end{figure}

\begin{table*}
\caption{\label{tab_results}Best fit results obtained for all sources included in the sample. }
\scriptsize 
\centering   
\begin{tabular}{lcccccccc} 
\hline
Source        & $N({\rm H}) $ - 21~cm &\NOi & \NOii  & \NOiii & \NOvi & \NOvii & \NOviii & \NOvi-{\it FUSE}\\
    & ($10^{21}$cm$^{-2}$) &  ($10^{16}$cm$^{-2}$) &  ($10^{16}$cm$^{-2}$) &  ($10^{16}$cm$^{-2}$) &  ($10^{16}$cm$^{-2}$) &  ($10^{16}$cm$^{-2}$) &  ($10^{16}$cm$^{-2}$) &  ($10^{16}$cm$^{-2}$)\\
    \\
\multicolumn{9}{c}{Galactic Sources  }\\
4U~0614+091& $4.42$& $ 144.97 \pm 0.22 $ & $ 5.41 \pm 1.28 $ & $ 3.28 \pm 0.67 $ & $< 0.64$  & $< 0.82$  & $< 0.22$ & -\\
4U~1636-53& $4.04$& $ 54.77 \pm 0.16 $ & $< 4.17$  & $< 6.42$  & $< 0.62$  & $ 1.57 \pm 0.50 $ & $ 1.02 \pm 0.25 $ &-\\
4U~1735-44& $3.96$& $ 75.41 \pm 0.50 $ & $< 13.77$  & $ 2.4 \pm 1.23 $ & $< 0.64$  & $< 2.06$  & $ 1.96 \pm 0.85 $& -\\
4U~1820-30& $2.33$& $ 81.06 \pm 0.10 $ & $ 3.48 \pm 0.89 $ & $ 1.10 \pm 0.25 $ & $< 0.20$  & $ 2.97 \pm 0.38 $ & $ 1.78 \pm 0.17 $ &-\\
Cygnus~X-2& $3.09$& $ 87.95 \pm 0.01 $ & $ 4.08 \pm 0.14 $ & $ 1.51 \pm 0.05 $ & $ 0.12 \pm 0.01 $ & $ 0.72 \pm 0.05 $ & $ 0.62 \pm 0.01 $&- \\
EXO~0748-676& $1.01$& $ 60.78 \pm 0.14 $ & $ 4.59 \pm 3.30 $ & $< 1.65$  & $< 3.99$  & $ < 1.17 $ & $< 0.77$&  -\\
GRO~J1655-40& $7.22$& $ 48.19 \pm 6.67 $ & $< 2.22$  & $< 0.90$  & $< 0.23$  & $< 1.25$  & $ 2.00 \pm 0.51 $&- \\
GX~339-4& $5.18$& $ 140.50 \pm 0.11 $ & $< 3.55$  & $ 1.08 \pm 0.25 $ & $< 0.43$  & $ 8.21 \pm 1.17 $ & $ 3.19 \pm 0.35 $& -\\
GX~349+2& $6.13$& $< 11.64$  & $< 1.73$  & $< 16.5$  & $< 0.33$  & $< 0.53$  & $ <3.33 $ &-\\
GX~9+9& $3.31$& $ 104.09 \pm 0.35 $ & $ 6.87 \pm 1.46 $ & $< 1.30$  & $< 0.69$  & $ 2.53 \pm 0.49 $ & $ 1.12 \pm 0.20 $& -\\
HER~X-1& $1.66$& $ 7.83 \pm 1.42 $ & $< 4.05$  & $< 0.82$  & $< 0.52$  & $ 4.06 \pm 2.90 $ & $< 0.27$ & -\\
SER~X-1& $5.42$& $ 41.37 \pm 0.12 $ & $< 3.09$  & $< 1.42$  & $< 0.35$  & $ 1.15 \pm 0.57 $ & $ 1.62 \pm 0.44 $&- \\
XTE~J1817-330& $2.29$& $ 78.52 \pm 0.01 $ & $ 8.75 \pm 0.03 $ & $ 1.23 \pm 0.01 $ & $ 0.20 \pm 0.02 $ & $ 1.80 \pm 0.02 $ & $ 1.20 \pm 0.02 $& -\\ 
\multicolumn{9}{c}{ Extragalactic Sources  }\\
1ES~0120+340&  $0.52$& $ 36.4 \pm 0.56 $ & $< 7.50$  & $< 0.82$  & $< 0.77$  & $< 0.73$  & $< 0.80$& - \\
1ES~1028+511&  $0.12$& $< 16.11$  & $< 5.31$  & $< 1.10$  & $< 1.97$  & $< 0.94$  & $< 2.59$& - \\
1ES~1553+113&  $0.43$& $ 13.21 \pm 0.31 $ & $ 1.91 \pm 0.42 $ & $< 0.44$  & $< 0.48$  & $ 0.36 \pm 0.07 $ & $< 1.01$&  -\\
1ES~1927+654&  $0.92$& $< 49.22$  & $< 6.83$  & $< 2.44$  & $< 7.78$  & $< 12.28$  & $< 1.64$ &- \\
1H0707-495&  $0.65$& $< 79.53$  & $< 55.05$  & $< 23.51$  & $< 1.78$  & $< 1.97$  & $< 3.47$ &- \\
3C~382&  $0.92$& $< 32.59$  & $< 16.23$  & $< 15.95$  & $< 10.67$  & $< 2.73$  & $< 7.59$ & $ 0.035^{+0.009}_{-0.007} $   \\
3C~273&  $0.17$& $ 12.31 \pm 0.35 $ & $ 2.20 \pm 0.37 $ & $< 0.61$  & $< 0.18$  & $ 1.64 \pm 0.14 $ & $ 0.51 \pm 0.05 $&$ 0.053^{+0.003}_{-0.002}	$ \\
3C~454.3&  $0.65$& $< 24.25$  & $ <10.35 $ & $< 2.53$  & $ <1.42 $ & $< 0.82$  & $ <3.11 $ &-\\
Ark~564&  $0.67$& $ 18.25 \pm 0.19 $ & $ 6.67 \pm 0.48 $ & $< 0.32$  & $< 0.12$  & $< 0.85$  & $< 0.53$&-  \\
B0502+675&  $1.45$& $< 3.29$  & $ <2.77 $ & $< 0.72$  & $ <1.90 $ & $ <3.29 $ & $< 2.49$& - \\
H1426+428&  $0.11$& $ 33.71 \pm 0.03 $ & $< 5.76$  & $< 6.29$  & $ <1.61 $ & $< 7.56$  & $< 3.81$& - \\
H1821+643&  $0.34$& $ 9.79 \pm 0.90 $ & $< 3.64$  & $< 2.04$  & $< 0.44$  & $< 0.54$  & $< 1.03$&$ 0.030\pm 0.003 $  \\
H2356-309&  $0.14$& $ 4.02 \pm 0.71 $ & $ 1.37 \pm 0.43 $ & $ 0.88 \pm 0.14 $ & $ 0.15 \pm 0.09 $ & $ 0.45 \pm 0.08 $ & $< 0.74$ &- \\
MCG-6-30-15&  $0.39$& $ 19.63 \pm 0.65 $ & $ 6.62 \pm 0.68 $ & $< 0.82$  & $< 0.68$  & $ 1.37 \pm 0.17 $ & $ 0.92 \pm 0.08 $ &-\\
MR~2251-178&  $0.26$& $< 15.33$  & $< 1.64$  & $< 5.62$  & $< 0.47$  & $< 2.79$  & $< 3.65$&-  \\
Mrk~279&  $0.17$& $ 21.57 \pm 0.21 $ & $ 3.32 \pm 0.38 $ & $ 1.09 \pm 0.13 $ & $< 0.10$  & $ 0.61 \pm 0.07 $ & $< 0.56$&$ 0.025\pm 0.001$  \\
Mrk~1044&  $0.38$& $< 25.18$  & $ 5.48 \pm 1.95 $ & $ 4.11 \pm 1.56 $ & $< 0.33$  & $< 3.07$  & $ <4.51 $&- \\
Mrk~421&  $0.20$& $ 4.93 \pm 0.10 $ & $ 1.66 \pm 0.11 $ & $< 0.82$  & $ 0.13 \pm 0.01 $ & $ 0.68 \pm 0.01 $ & $ 0.15 \pm 0.02 $& $ 0.030_{-0.009}^{+0.005}$ \\
Mrk~290&  $0.17$& $ 11.88 \pm 3.74 $ & $<1.47 $ & $< 0.79 $ & $< 0.26$  & $ 2.08 \pm 1.05 $ & $ <0.71 $&$ 0.016\pm 0.006$ \\
Mrk~509&  $0.50$& $ 20.02 \pm 0.21 $ & $< 4.77$  & $ 1.86 \pm 0.22 $ & $< 0.49$  & $ 1.21 \pm 0.36 $ & $< 1.02$ &$0.045\pm 0.003$ \\
NGC~3783&  $1.38$& $ 130.71 \pm 1.78 $ & $< 5.67$  & $ <2.01 $ & $ <0.88 $ & $ <1.12 $ & $ 0.53 \pm 0.42 $&- \\
NGC~4051&  $0.11$& $ 4.64 \pm 0.68 $ & $ 2.56 \pm 0.45 $ & $< 1.78$  & $ <0.63$ & $ < 1.51 $ & $ < 3.38 $& -\\
NGC~4593&  $0.20$& $ 5.30 \pm 1.60 $ & $< 3.63$  & $< 1.02$  & $< 0.65$  & $ 0.86 \pm 0.35 $ & $< 2.22$ & -\\
NGC~5548&  $0.16$& $ 6.61 \pm 0.74 $ & $ 2.10 \pm 0.72 $ & $< 1.44$  & $< 0.07$  & $< 0.75$  & $ 1.36 \pm 0.27 $ &$ 0.031\pm 0.004 $\\
NGC~7469&  $0.52$& $ 42.86 \pm 1.70 $ & $< 29.26$  & $< 3.41$  & $< 0.24$  & $< 4.70$  & $< 4.91$ &$ 0.0091_{-0.0012}^{+0.0009} $ \\
PG~1211+143&  $0.30$& $ 36.25 \pm 0.07 $ & $< 6.33$  & $< 6.01$  & $< 1.62$  & $ 0.92 \pm 0.61 $ & $< 2.33$ & $ 0.014\pm 0.002 $\\
PKS~2005-489&  $0.46$& $ 8.67 \pm 2.58 $ & $< 10.26$  & $< 3.31$  & $< 2.21$  & $< 0.65$  & $< 0.74$ &$ 0.060\pm 0.003 $ \\
PKS~2155-304&  $0.16$& $ 10.99 \pm 0.37 $ & $ 2.63 \pm 0.20 $ & $< 0.19$  & $< 0.08$  & $ 0.47 \pm 0.02 $ & $ 0.46 \pm 0.03 $&$0.0218\pm 0.0005 $ \\
Ton~1388&  $0.15$& $< 21.59$  & $ 10.81 \pm 3.25 $ & $< 2.38$  & $< 0.31$  & $ <2.72 $ & $<1.59 $ &$0.018_{-0.002}^{+0.001} $\\
\hline
\multicolumn{9}{p{15cm}}{ 21~cm measurements are taken from \citet{wil13}. {\it FUSE} measurements are taken from \citet{sav03}.}

\end{tabular}
\end{table*} 
 
\section{Results and discussion}\label{sec_results}

Table~\ref{tab_results} shows the oxygen column densities obtained for all
sources analyzed. \NOvi\ obtained by \citet{sav03} using observations from the Far Ultraviolet Spectroscopic Explorer ({\it
FUSE})  are included for those
sources for which data is available.  We have found that both \NOii\ and \NOvi\ can be simultaneously constrained only for 4 sources in the
sample: Cygnus~X--2, XTE~J1817-330, H2356--309 and Mrk~421.  For the rest of the sources we have measured upper limits for
the \NOvi. We also have measured the column densities for each observation of the sources (i.e. instead of combining the spectra) and we found that the \NOvi\ do not vary between different observations (considering the uncertainties), pointing out to a ISM origin of the absorber.  For Mrk~421, Mrk~509 and PKS~2155-304 our results are consistent with \citet{mat17}.  Because we are modeling the complete oxygen edge
absorption region the maximum amount of \NOii\ is constrained by the presence
of both the K$\alpha$ and the K$\beta$ transitions.  Moreover, the contribution
of the photoabsorption cross-section from the multiple ions to the continuum is
not negligible. From Table~\ref{tab_results}, it is clear that there is a discrepancy with X-ray best-fit results and the UV values obtained from {\it FUSE} observations by \citet{sav03}. That is, even when
the \NOii\ is accurately modeled, the \NOvi\ estimated from X-ray observations can be up to two orders of
magnitude larger than the UV measurements.  Using the column densities derived with the {\tt IGMabs} model we have computed the EWs for the \Ovi\ K$\alpha$ and \Oii\ K$\beta$-L12 absorption lines. Figure~\ref{fig3} shows the percentage contribution of \Oii\ K$\beta$-L12 to the total (\Ovi\ K$\alpha$ + \Oii\ K$\beta$-L12) EW. In most cases the \Oii\ K$\beta$-L12 accounts for <$30\%$ of the total EW. Such contribution is not enough to solve the \NOii\ and \NOvi\ discrepancy. In the cases of Cygnus~X--2 and XTE~J1817--330, the \Oii\ K$\beta$-L12 accounts for $\sim 40\%$ of the total EW.
 
Figure~\ref{fig_mrk421} shows the 21--24~\AA\ wavelength region for Mrk~421, a source for which we have obtained a good constraint for both \NOii\ and \NOvi. The most
outstanding features associated with oxygen X-ray absorption in the ISM are
also indicated. We tried to model the absorption line located at $\sim 22.04$ \AA\ by fixing the \NOvi\ in the \IGMabs\ model to the {\it FUSE} measurement and increasing the \NOii\ until we obtain low residuals around the line wavelength position.  We found that this resulted in worse fit statistics because when assuming the {\it FUSE} \NOvi\ a larger amount of \Oii\ is required to model the $\sim 22.04$ \AA\ line, overestimating the strengths of other lines, in particular the \Oii\ K$\alpha$, located at $23.35$ \AA.  Figure~\ref{fig_mrk421} also included the data/model ratios obtained for those sources for which {\it FUSE} measurements are available when applying the same procedure. As before, the residuals around the \Oii\ K$\alpha$ absorption line increase as we increase the \NOii. In all cases the statistics is worse by $\Delta\chi^{2}>15$, except for NGC~7469 ($\Delta\chi^{2}=8$) and Ton~1388 ($\Delta\chi^{2}=6$). These results indicate
that the blending of \Ovi\ K$\alpha$ and the \Oii\ K$\beta$-L12 absorption features
it is not sufficient to account for the discrepancy between X-ray and UV measurements.
 
Other possibilities to explain the \Ovi\ mismatch between UV and X-ray measurements include the atomic data calculation, saturation of the lines and radiative excitation.  In the case of \Oii, given that the Auger widths tend to be larger than the radiative widths \citep{gar05} it is crucial to include such effect, which can lead to an increasing of the EW up to $50\%$ for a fixed column density. However, it is not trivial to compute all the multiple channels for the Auger decay of the K vacancy states, making difficult to estimate accurately the Auger damping. Also, as was noted by \citet{biz15}, there is a discrepancy between the theoretical and experimental values for the oscillator strengths and line width for \Oii\ K$\beta$-L12, which has not been addressed.  In the case of saturated lines, given the dependence of the EW with the broadening velocity $v_{turb}$,  there is a \NOvi-$v_{turb}$  degeneracy which can lead to under(over)estimation of the column density  \citep[up to one of magnitude for $v_{turb}>200$ km~s$^{-1}$, see][]{dra11,nic16a}. UV data provides a good constraint on the velocity because multiple transitions of the same ion can be modeled simultaneously, thus allowing the disentangling of such degeneracy \citep{sav06,bow08}. However, even when the $v_{turb}$ value assumed in our model is within the broadening velocity found in UV studies \citep[see for example][]{sav06} the mismatch between the column densities remains.  With respect to radiative excitation, as mentioned by \citet{ara03}, electron impact excitation and recombination may contribute to the \Ovi\ emission, reducing the absorption from the ground state. But such effect can only account for one-half factor of difference between UV and X-ray, much lower than the discrepancy observed \citep{mat17}.

   \begin{figure}
   \begin{center}
     \includegraphics[scale=0.45]{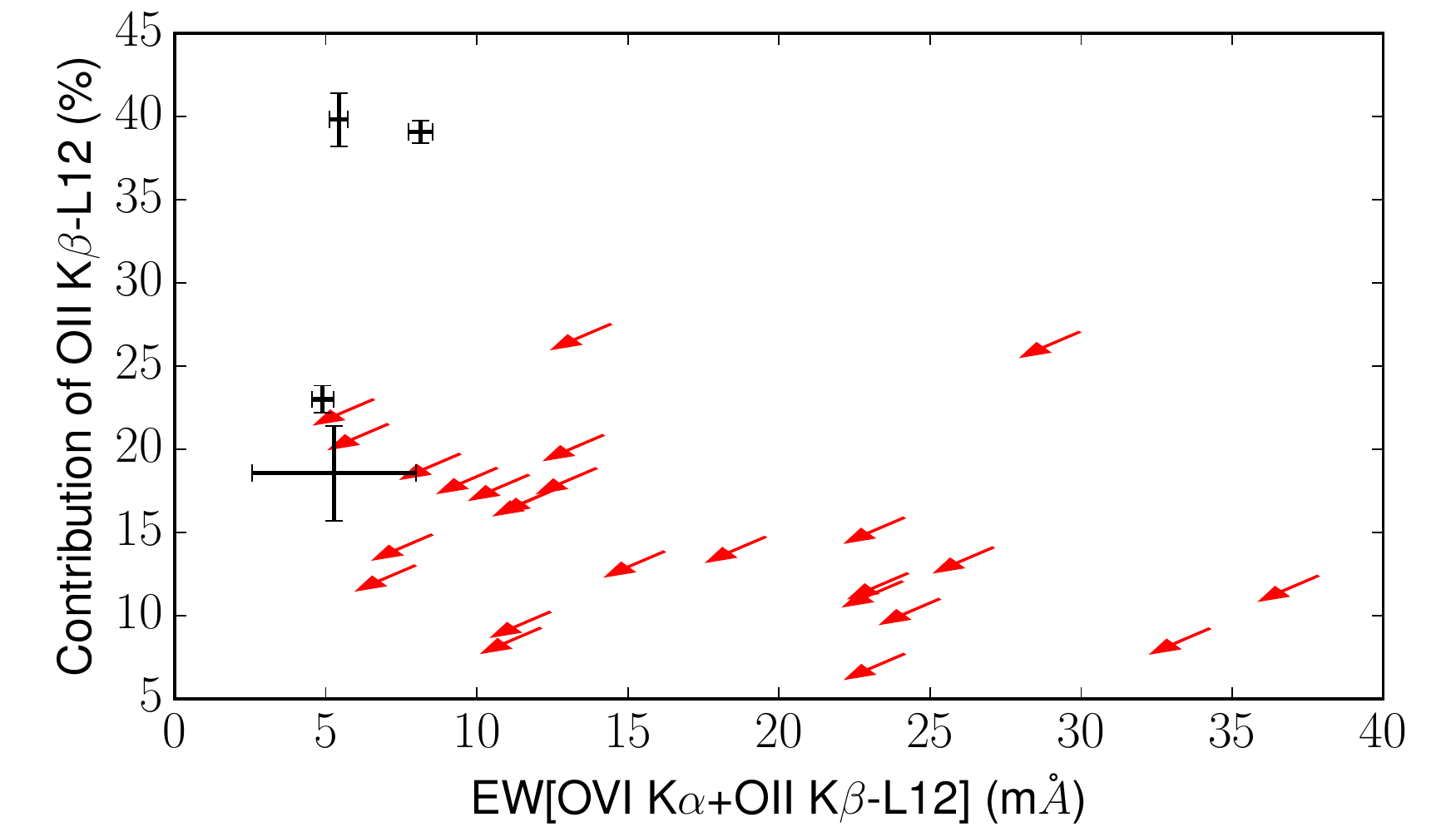}
     \vspace*{-1mm}
     \caption{Percentage contribution of the \Oii\ K$\beta$-L12 line to the total (\Oii\ K$\beta$-L12+\Ovi\ K$\alpha$) EW. Black points are for sources where both \Oii\ K$\beta$-L12 and \Ovi\ K$\alpha$ are measured. Red arrows are for sources where the two lines have only upper limits.}\label{fig3}
     \end{center}
   \end{figure}

     \begin{figure}
   \begin{center}
     \includegraphics[scale=0.425]{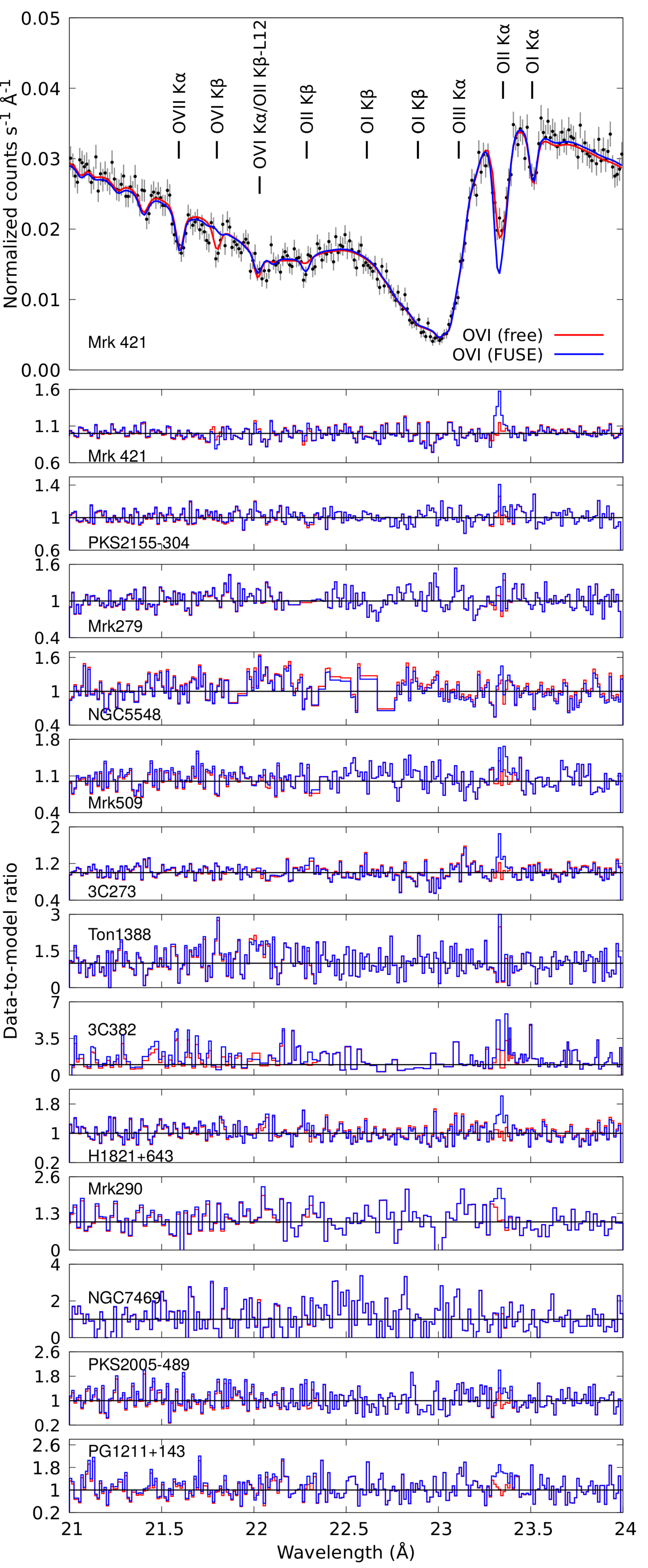}
          \vspace*{-1mm}
     \caption{Oxygen edge absorption region for those sources in the sample that have both X-ray and UV column density measurements. Red solid lines correspond to the best fit model obtained when \NOvi\ is a free parameter (see Table~\ref{tab_results}). Blue solid line correspond to the fit obtained by fixing  \NOvi\ to  {\it FUSE} value and increasing \NOii\ until model the $\sim 22.04$ \AA\ absorption line.
     }\label{fig_mrk421}
     \end{center}
   \end{figure} 
  
\section{Conclusions}\label{sec_con}
 
We have analyzed the oxygen edge absorption region (18--24~\AA) using
high-resolution X-ray spectra from 13 LMXBs and 29 extragalactic sources. Using the \IGMabs\ model we have estimated column
densities, ion by ion, required to fit all the absorption signatures in the
spectra. We have focused the analysis in the estimation of \NOvi,
which can be difficult due to contamination by the \Oii\ K$\beta$-L12 absorption
feature at $22.04$ \AA\, located close to the \Ovi\ K$\alpha$ absorption
line at $22.032$ \AA. \NOvi\ was constrained for 2
Galactic and 2 extragalactic sources from our sample. For the rest of the
sources we report upper limits. We have found that for low \NOii, the \NOvi\ contribution is larger, a hint of the blending between
the line. However, this blending cannot be the entire solution for the
discrepancy between X-ray and UV measurements of \NOvi. Other possibilities include atomic data calculation, saturation of the lines and radiative excitation, however they cannot completely solve the UV/X-ray discrepancy neither, which remains as an open question. Finally, our findings highlight the need for an accurate modeling of the absorption features associated to the local ISM before searching for lines associated to other absorbers, including the warm-hot intergalactic medium, ultraluminous X-ray sources, active galactic nucleus and x-ray binaries, which constitute key science goals for future missions such as {\it Arcus} and {\it Athena}.

\section{Acknowledgements}
 This research was supported by the DFG cluster of excellence `Origin and Structure of the Universe'. E.G. thank Amit Pathak, 
Ken Sembach and Bart Wakker for helpful comments on the UV absorption line measurements. J.A.G. acknowledges support from NASA grant 80NSSC17K0345 and from the Alexander von Humboldt Foundation.

\bibliographystyle{mnras}

\end{document}